# Rytov Approximation of Vector Waves by Modifying Scattering Matrices: Precise Reconstruction of Dielectric Tensor Tomography

ChulMin Oh,[1,2] Herve Hugonnet,[1,2] Juheon Lee,[1,2] and YongKeun Park[1,2,3,*]

[1]*Department of Physics, Korea Advanced Institute of Science and Technology, Daejeon 34141, Republic of Korea.*
[2]*KAIST Institute for Health Science and Technology, Korea Advanced Institute of Science and Technology, Daejeon, 34141, Republic of Korea*
[3]*Tomocube Inc., Daejeon 34051, Republic of Korea*
[*]*yk.park@kaist.ac.kr*

Analyzing 3D anisotropic materials presents significant challenges, especially when assessing 3D orientations, material distributions, and anisotropies through scattered light, due to the inherent vector nature of light-matter interactions. In this study, we formulate a scattering theory based on the Rytov approximation, commonly employed in scalar wave tomography, tailored to accommodate vector waves by modifying the scattering matrix. Using this formulation, we investigate the intricate 3D structure of liquid crystals with multiple topological defects exploiting dielectric tensor tomography. By leveraging dielectric tensor tomography, we successfully visualize these topological defects in three dimensions, a task that conventional 2D imaging techniques fail to achieve.

## I. INTRODUCTION

The molecular orientation of anisotropic materials in three-dimensional space is crucial across numerous applications, from advanced material design to biological understanding [1-3]. Traditionally, polarization microscopy has served as the primary tool for qualitatively assessing their orientational order in 2D. Recently, several microscopy techniques exploiting nonlinear optical effects have been used to image the 3D orientations [4-6]. Although these techniques offer qualitative insights into director distributions, it remains a challenge to accurately quantify the organization of anisotropic materials and the degree of their alignment. Overcoming this challenge requires direct measurements of the dielectric permittivity tensor, which incorporates direction ($\hat{n}_e$), refractive indices ($n_e$, $n_o$), and anisotropy ($\Delta n$) [1].

Tomographic reconstruction of scattering properties such as the dielectric tensor can be achieved by using the information of the scattered waves under known illuminations, which involves solving and inverting the wave equation [7,8]. Given the complexity of solving the wave equation, most of existing tomographic methods approximate solutions. While the first-order Born approximation commonly encountered in physics is one option, it is not particularly favored for reconstructing tomograms of thick objects due to the slow convergence of the Born series. Instead, the Rytov series offers much faster convergence by expanding the logarithm of the field into a series rather than the field itself. By using a nonlinear function, the first-order Rytov approximation can include higher order terms than the Born series [9,10].

In scalar scattering theory, the Rytov series can be derived straightforwardly by comparing it with the Born series [11-13]. However, extending this to vector scattering presents a challenge: replacing the logarithm of scalar fields with that of tensor fields. Since light is a transverse wave, at most two orthogonal polarizations can correspond to a given incident plane wave. Consequently, the logarithm of tensor fields composed of two vector fields always diverges due to their zero eigenvalue [7]. Previous studies have struggled to solve this problem, with some proposing incomplete Rytov approximations by applying the scalar Rytov approximation to each component of the vector field [8,14,15], or introducing a paraxial approximation that ignores components of the vector waves parallel to the propagation direction [7].

In this study, we address this issue by modifying the scattering matrix before extending it into the Rytov series. We present an efficient method for inverting the Rytov approximation and demonstrate precise and accurate reconstruction of dielectric tensor tomography. The method is experimentally validated using measurements of complex liquid crystal networks containing several 3D topological defects.

## II. SCATTERING THOERY

Elastic scattering can be described by the wave equation [16,17]:

$$(E - H_0)\psi = V\psi, \qquad (1)$$

where $H_0$ is an unperturbed Hamiltonian, $E$ is the unperturbed energy of the incident wave, and $V$ is the scattering potential of an object, which act as a perturbing Hamiltonian. In nonmagnetic and dielectric linear media, the precise form of $H_0$, $E$, and $V$ can be derived from the inhomogeneous Maxwell's equations with $\psi$ in Eq. (1) being the electric field [9]:

$$\begin{aligned}
H_0 &= \nabla \times \nabla \times, \\
E &= k_0^2, \\
V &= -k_0^2 \left( \boldsymbol{\varepsilon}_\mathbf{r} / n_m^2 - \mathbf{1} \right),
\end{aligned} \qquad (2)$$

where $k_0$ is the wavevector of the incident light, $\boldsymbol{\varepsilon}_\mathbf{r}(\mathbf{r})$ is the dielectric permittivity tensor of the object, and $\mathbf{1}$ is an identity tensor.

To relate the incoming (outgoing) wave in the free space $\phi_{in}$ ($\phi_{out}$) and the corresponding solution of the wave equation $\psi^+$ ($\psi^-$), the Lippmann–Schwinger equation is introduced as follows [16,17]:

$$\psi^\pm = \phi_{in/out} + G_{0\pm} V \psi^\pm, \quad (3)$$

$$G_{0\pm} = \frac{1}{E - H_0 \pm i0^+}, \quad (4)$$

where $G_{0\pm}$ is a resolvent (Green's) operator.

Thus, the simple linear relation $\psi^\pm = \Omega_\pm(E)\phi_{in/out}$ can be established using the Møller matrix [16]:

$$\Omega_\pm(E) = (1 - G_{0\pm} V)^{-1}. \quad (5)$$

To formulate the Rytov approximation for vector waves, one may try to expand the logarithm of Eq. (5) into a series. To keep each term of the series a tensor, its matrix logarithm must be used. In other words, the logarithm must be applied to its eigenvalues rather than its component. However, due to the transverse nature of light, Møller matrix has only two physically meaningful (observable) eigenvalues corresponding to two incident polarizations, which poses a problem when applying the logarithm. To address this challenge, we introduce an extended Møller matrix by adding a non-interacting scattering channel orthogonal to the polarization subspace:

$$\mathbf{M}_{\Omega_+,\mathbf{k}}(\mathbf{r}) = \langle \mathbf{r}|\Omega_+|\mathbf{k}\rangle P_{\mathbf{k}\perp} + \langle \mathbf{r}|\mathbf{k}\rangle P_{\mathbf{k}\parallel}, \quad (6)$$

where $P_{\mathbf{k}\parallel} = \hat{\mathbf{k}}\hat{\mathbf{k}}$ and $P_{\mathbf{k}\perp} = \mathbf{1} - P_{\mathbf{k}\parallel}$ are projection operators onto polarization subspaces parallel and orthogonal to $\mathbf{k}$, respectively, and $|\mathbf{k}\rangle$ corresponds to an incident plane wave whose position-domain representation is $\langle \mathbf{r}|\mathbf{k}\rangle = e^{i\mathbf{k}\cdot\mathbf{r}}$. With the extended Møller matrix now having three well-defined and nonzero eigenvalues, its Rytov series,

$$\mathbf{M}_{\Omega_+,\mathbf{k}}(\mathbf{r}) = \langle \mathbf{r}|\mathbf{k}\rangle \exp\left(\sum_{m=0}^{\infty} \mathbf{\Psi}_m\right), \quad (7)$$

can be derived from the Born series:

$$\mathbf{M}_{\Omega_+,\mathbf{k}}(\mathbf{r}) = \langle \mathbf{r}|\mathbf{k}\rangle \mathbf{1} + \langle \mathbf{r}|G_{0+}V|\mathbf{k}\rangle P_{\mathbf{k}\perp} + \langle \mathbf{r}|(G_{0+}V)^2|\mathbf{k}\rangle P_{\mathbf{k}\perp} + \cdots. \quad (8)$$

where the exact form of $\mathbf{\Psi}_m$,

$$\mathbf{\Psi}_0 = 0,$$

$$\mathbf{\Psi}_1 = \frac{\langle \mathbf{r}|G_{0+}V|\mathbf{k}\rangle}{\langle \mathbf{r}|\mathbf{k}\rangle} P_{\mathbf{k}\perp},$$

$$\mathbf{\Psi}_2 = \frac{\langle \mathbf{r}|(G_{0+}V)^2|\mathbf{k}\rangle}{\langle \mathbf{r}|\mathbf{k}\rangle} P_{\mathbf{k}\perp} - \frac{1}{2!}\mathbf{\Psi}_1^2,$$

$$\mathbf{\Psi}_3 = \frac{\langle \mathbf{r}|(G_{0+}V)^3|\mathbf{k}\rangle}{\langle \mathbf{r}|\mathbf{k}\rangle} P_{\mathbf{k}\perp} - \frac{1}{2!}(\mathbf{\Psi}_1 \mathbf{\Psi}_2 + \mathbf{\Psi}_2 \mathbf{\Psi}_1) - \frac{1}{3!}\mathbf{\Psi}_1^3, \quad (9)$$

can be obtained by equating Eqs. (7) and (8) so that $\mathbf{\Psi}_m$ becomes the order of $(G_{0+}V)^m$. Note that $\mathbf{\Psi}_1 \mathbf{\Psi}_2 \neq \mathbf{\Psi}_2 \mathbf{\Psi}_1$ since they are tensors.

To test the derived formulas, we calculated the waves scattered by a radially oriented nematic liquid crystal droplet [Fig. 1(a)]. The droplet has a diameter of 6 μm, with the refractive index (RI) of the medium set to 1.5 and those of the liquid crystal to 1.56 for extraordinary light and 1.5 for ordinary light. The incident light is an x-polarized plane wave with a wavelength of 532 nm traveling along the z-direction. Figure 2(b) displays root-mean-square error of Born and Rytov series:

$$\sqrt{\left\langle \left|\psi_{Born/Rytov} - \psi_{exact}\right|^2 / \left|\phi_{in}\right|^2 \right\rangle} \quad (10)$$

where $\psi_{exact}$ is the exact solution of the Lippmann–Schwinger equation [Eq. (3)]. Notably, not only does the Rytov approximation have much less error than the Born approximation of the same order, but the first-order Rytov approximation also outperforms the third-order Born approximation. Figure 1(c) displays the x-component of the normalized scattered wave $\psi_x/\phi_{\text{in},x}$. The remaining components are shown in Supplemental Material. While the Rytov approximation generates phase and amplitude similar to the exact solution, the Born approximation deviates significantly from the exact solution due to the large RI of the object. In particular, the Born approximation results in small phase values, leading to low RIs when applied to tomography.

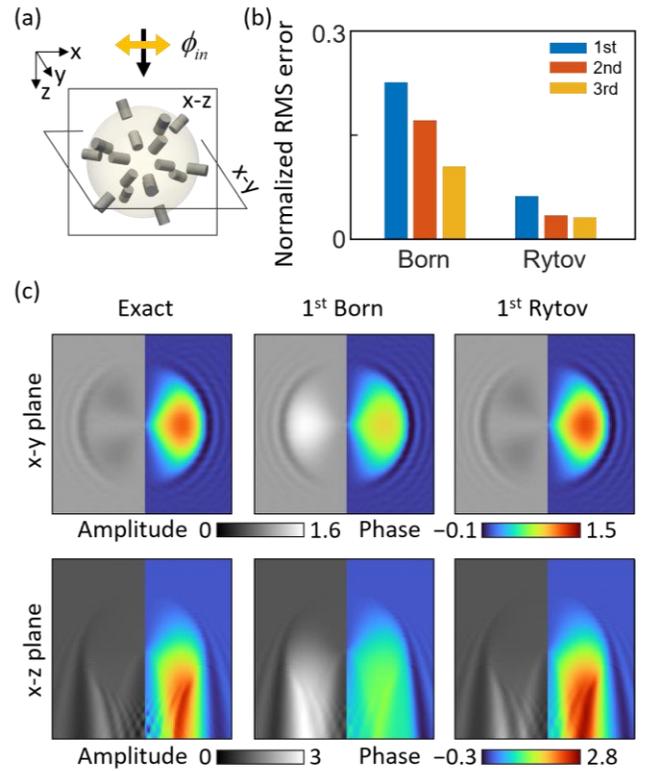

FIG. 1. Calculation of electric fields scattered by a radially oriented nematic liquid crystal droplet. (a) Light of 532 nm wavelength is incident on a 6-μm-diameter bead of radially oriented nematic liquid crystal. The incident electric field is polarized in the x direction at a wavelength of 532 nm. The refractive index (RI) of the medium is 1.5 and that of the liquid crystal is 1.56 for extraordinary light and 1.5 for ordinary light. (b) Normalized root-mean-square (RMS) error of the electric fields calculated using Born and Rytov series. (c) x-component of the calculated electric fields next to the exact solution of the Lippmann–Schwinger equation.

In experimental situations, what can be measured are waves in free space $\phi_{\text{in}}$ and $\phi_{\text{out}}$, rather than the actual solution $\psi$. Thus, the scattering matrix, $S = \Omega_-^\dagger \Omega_+$, which directly connects the incoming and outcoming waves as $\phi_{\text{out}} = S\phi_{\text{in}}$, is a more useful quantity for tomographic reconstruction [16]. To derive the Rytov approximation for the scattering matrix, we again introduce the following modification:

$$\mathbf{M}_{S,\mathbf{k}}(\mathbf{r}) = \langle \mathbf{r}|S|\mathbf{k}\rangle P_{\mathbf{k}\perp} + \langle \mathbf{r}|\mathbf{k}\rangle P_{\mathbf{k}\parallel}. \quad (11)$$

Since the scattering matrix can be directly related to $\Omega_+$ via following relation [16],

$$\langle \mathbf{k}'|S|\mathbf{k}\rangle = \langle \mathbf{k}'|\mathbf{1} + G_{\text{far}}V\Omega_+|\mathbf{k}\rangle, \quad (12)$$

where we used the far-field asymptote of $G_{0\pm}$, which satisfies

$$\langle \mathbf{k}'|G_{\text{far}}|\mathbf{k}\rangle = -2\pi i \delta(k^2 - k_0^2) P_{\mathbf{k}\perp} \langle \mathbf{k}'|\mathbf{k}\rangle. \quad (13)$$

The Born and Rytov series of the extended scattering matrix can be simply obtained by replacing the leftmost $G_{0\pm}$ in those of the extended Møller matrix with $G_{\text{far}}$ (see Supplemental Material for details). For example, the first order Rytov approximation becomes

$$\mathbf{M}_{S,\mathbf{k}}(\mathbf{r}) \approx \langle \mathbf{r}|\mathbf{k}\rangle \exp\left(\langle \mathbf{r}|\mathbf{k}\rangle^{-1} \langle \mathbf{r}|G_{\text{far}}V|\mathbf{k}\rangle\right) P_{\mathbf{k}\perp}. \quad (14)$$

## III. TOMOGRAPHIC INVERSION METHOD

To reconstruct the dielectric tensor tomogram of an object from the measured scattering matrix using the Rytov approximation, we derived the inversion formula of the Eq. (14):

$$\delta(k'^2 - k_0^2) P_{\mathbf{k}'\perp} \tilde{V}(\mathbf{k}' - \mathbf{k}) P_{\mathbf{k}\perp}$$
$$= \frac{i}{2\pi} \int \log\left(\mathbf{M}_{S,\mathbf{k}}(\mathbf{r}) e^{-i\mathbf{k}\cdot\mathbf{r}}\right) e^{-i(\mathbf{k}'-\mathbf{k})\cdot\mathbf{r}} d\mathbf{r}, \quad (15)$$

where $\tilde{V}(\mathbf{k}' - \mathbf{k}) = \int V(\mathbf{r}) e^{-i(\mathbf{k}'-\mathbf{k})\cdot\mathbf{r}} d\mathbf{r}$. Note that computing the logarithm of tensors is not a trivial problem due to the ambiguity of the complex logarithm. We have addressed this problem by developing a phase unwrapping method for tensors (see Supplemental Material for details).

Using Eq. (15), the scattering potential on the Ewald sphere (represented by the delta function) can be revealed from the measured scattering matrix. Assuming symmetry of the dielectric tensor [18], there are 6 unknown components of the scattering potential. However, Eq. (15) yields only 4 independent linear equations due to the two projection operators, indicating that the inversion problem is underdetermined. We address this problem by focusing on the intersections of the Ewald spheres corresponding to different incident wave vectors, where there are 7 independent equations for 6 unknown variables.

To reduce the computational complexity of the inversion process, we minimize the error function, rather than directly solving the linear equation at the intersection points. The error function incorporates the mean squared error between each side of Eq. (15), and the regularization term:

$$\left\|V - \mathbf{1}\text{tr}(V/3)\right\|^2_{\text{Frobenius norm}}, \quad (16)$$

which suppresses false anisotropy signals from nonintersecting points (see Supplemental Material for the detailed description on error function).

## IV. EXPERIMENTAL RESULTS

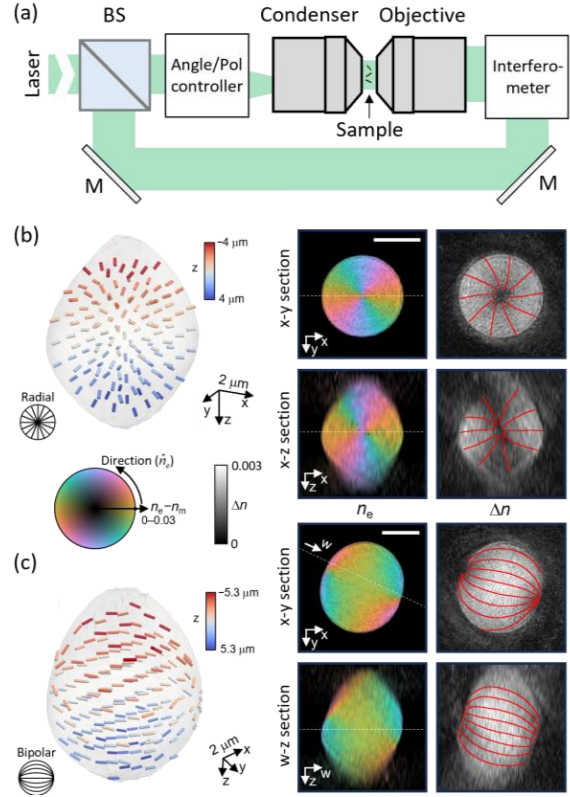

FIG. 2. Tomography of liquid crystal droplets. (a) Experimental setup for measurement of scattering (transmission) matrices. BS: beamsplitter. M: mirror. (b-c) Director field, RI, and birefringence of liquid crystal droplets in (b) radial and (c) bipolar configuration. Red lines: the 2D streamlines of the director fields. The RI of the medium ($n_m$) is 1.5247 for the laser used. Scale bars in (b) and (c) are 5 μm.

To demonstrate the proposed method, we reconstruct dielectric tensor tomograms from the scattering matrices measured using a setup illustrated in Fig. 2(a). In this setup, plane waves from a laser with a wavelength of 532 nm are transmitted to the sample through a condenser lens (UPLSAPO 60XW, ×60, Olympus) at different incident angles and polarizations. Then, the transmitted waves are collected by an objective lens (UPLSAPO 60XO, ×60, Olympus) and recorded by an off-axis holography interferometer in two polarization bases. Detailed optical elements of the setup are provided in the Supplemental Material, together with references [8,15].

Figures 2(a) and (b) display the reconstructed tomogram of two liquid crystal droplets in radial and bipolar configurations, prepared by surfactants of sodium dodecyl sulfate and polyvinyl alcohol, respectively [19-21]. Since the liquid crystal, made of reactive mesogen mixture, is positive uniaxial, we identify the square-root of the largest eigenvalue of dielectric tensor as the extraordinary index ($n_e$), and the corresponding eigenvector as the director field ($\hat{n}_e$). The square-root of the smallest eigenvalue is taken as the ordinary index ($n_o$). The birefringence, implying anisotropy, is then calculated as the difference between the indices ($\Delta n = n_e - n_o$).

Reconstructed 3D director fields follow known directions in both configurations. The topological defects in radial and bipolar configuration can be observed not only from the director fields but also from the birefringence since the macroscopic birefringence is cancelled out in these defects. Therefore, the spatial distribution of the liquid crystal and how well it is ordered can be estimated by considering both the refractive index and birefringence simultaneously. The stretching of tomograms in the axial direction is due to the limited range of illumination angles, commonly referred to as the missing cone problem in optical tomography [22].

We then reconstructed tomograms of a liquid crystal network film patterned by an electric field applied to monomeric mixtures [23]. The reconstructed director field, shown in Figure 3(a), reveals horizontal and vertical bands composing the network, as well as upright directors between the bands and topological defects at the center and intersection of the bands. This result is consistent with the structure estimated by a commercial liquid crystal simulator (TechWiz LCD 3D, Sanayi System Co., Ltd.) based on the applied electric field and material properties [Fig. 3(b)]. It is important to note that the simulated result does not predict the actual director field, but rather its tendency, since the exact electric field inside the materials is inaccessible. In addition, Figure 3(c) shows the strong birefringence of the network, and weak birefringence near the topological defects.

To investigate the topological defects from the reconstructed tomogram, we plot the director fields near the defects (i)–(iii) in Fig. 3(d). The defects (i) and (ii) are identified as radial boojums oriented to +z and −z, respectively, which are difficult to distinguish in 2D imaging. Figure 3(d) also clearly visualizes the 3D structure of a defect (iii): as a horizontal band crosses over a vertical band, a topological defect of charge $m = -1$ is formed in between.

For comparison, the tomograms reconstructed using the first order Born approximation and the component-wise scalar Rytov approximation [8] are shown in Fig. 4. The directors from the Born approximation follow a similar trend to the simulation result but show irregular changes and are noisier due to the low and inaccurately acquired refractive index and birefringence. On the other hand, neither the directors nor the refractive indices from the scalar Rytov approximation provide meaningful information, implying that it can only be applied to samples with negligibly weak birefringence [24].

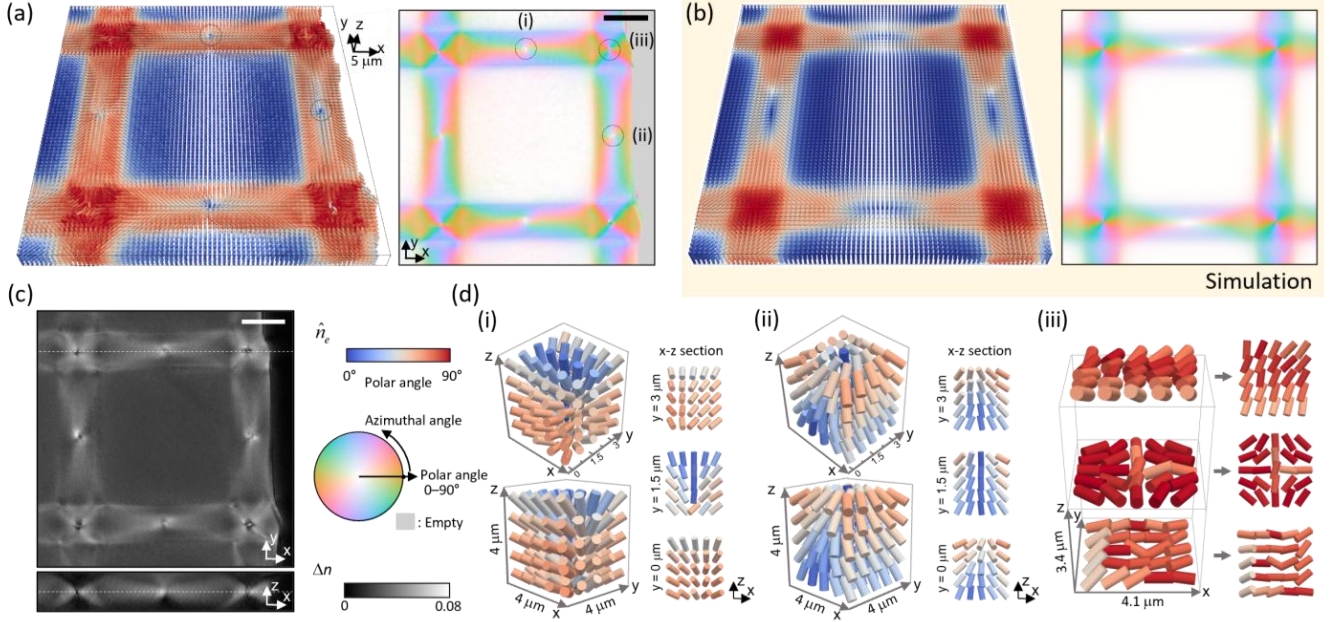

FIG. 3. Tomography of a liquid crystal network film. (a) Measured director fields. (b) Director fields estimated by a commercial liquid crystal simulator based on the design parameters. (c) 3D tomogram of birefringence ($\Delta n$). (d) Zoomed view of the director fields near the selected topological defects. Scale bars in (a) and (c) are 10 μm.

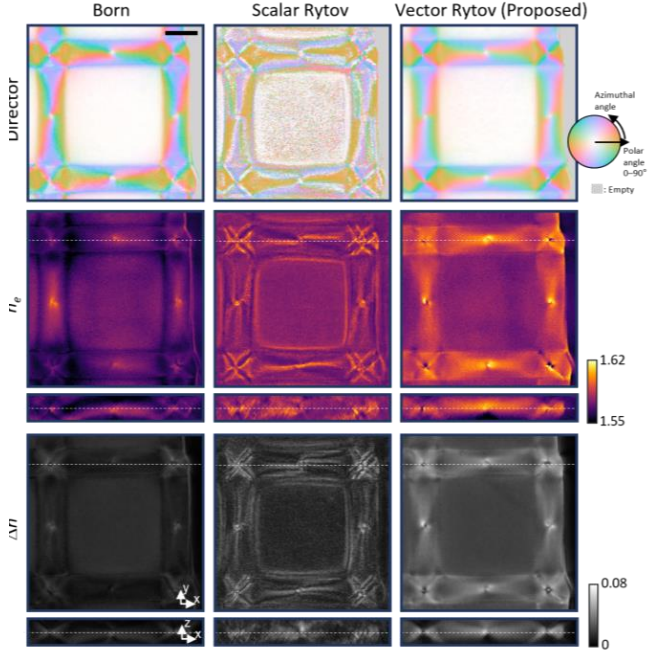

FIG. 4. Comparison of tomograms reconstructed using the Born approximation, component-wise scalar Rytov approximation, and proposed vector Rytov approximation. Scale bar is 10 μm.

## V. CONCLUSION

In this study, we extended the Rytov approximation to embrace the vector nature of electromagnetic waves, introducing a method for the tomographic reconstruction of the dielectric permittivity tensor in three dimensions. This enabled us to reconstruct the director fields, refractive indices, and birefringence of a strongly birefringent liquid crystal network with great details, providing new insights into the complex structure of 3D topological defects.

It holds the potential for wide-ranging applications, from advancing material science to enhancing our comprehension of optical phenomena involving the diffraction of vector fields [25]. Beyond electromagnetic waves, our formulation can be applied to general scattering phenomena. In general, the speed of a wave (and thus the magnitude of the wave vector) depends on its polarization [26-29]. Thus, the following extended scattering matrix can be constructed as before:

$$\mathbf{M}_{S,\mathbf{k}}(\mathbf{r}) = \langle \mathbf{r} | S | \mathbf{k} \rangle P_{\mathbf{k}} + \langle \mathbf{r} | \mathbf{k} \rangle (\mathbf{1} - P_{\mathbf{k}}) \quad (17)$$

where $\mathbf{k}$ is the wave vector of the incoming plane wave and $P_{\mathbf{k}}$ is a projection operator onto the polarization subspace corresponding to $\mathbf{k}$. We therefore expect that our formulation will also be useful for studying the scattering of mechanical waves such as elastic (or seismic) waves [30-35], the main source of imaging in geophysics [36-39].

The supporting data and codes for this article are openly available from GitHub [40]

## ACKNOWLEDGMENTS

This work was supported by National Research Foundation of Korea (2015R1A3A2066550, 2022M3H4A1A02074314), Institute of Information & communications Technology Planning & Evaluation (IITP; 2021-0-00745) grant funded by the Korea government (MSIT), KAIST Institute of Technology Value Creation, Industry Liaison Center (G-CORE Project) grant funded by MSIT (N10240002).